\documentclass[doublecol,figures]{epl2}
\usepackage{graphicx,amssymb,amsmath,setspace,stmaryrd,bm}

\title{Asymptotics for the survival probability of a Rouse chain monomer 
}
\shorttitle{Title}

\author{
G. Oshanin\thanks{E-mail: \email{oshanin@lptl.jussieu.fr}}
}

\shortauthor{G. Oshanin}

\institute{
Laboratoire de Physique Th{\'e}orique de la Mati{\`e}re
Condens{\'e}e, Universit{\'e} Pierre et Marie Curie (Paris 6) -
4 Place Jussieu, 75252 Paris, France
}

\pacs{05.40.Fb}{Random walks and Levy flights}
\pacs{02.50.Ey}{Stochastic processes}

\abstract{We study the long-time asymptotical behavior of the survival probability $P_t$ 
of a tagged monomer of an infinitely long Rouse chain in presence of 
two fixed absorbing boundaries, placed at $x = \pm L$.
Mean-square displacement of a tagged monomer obeys $\overline{X^2(t)} \sim t^{1/2}$ at all times,
which signifies that its dynamics 
is an anomalous diffusion process. 
Constructing lower and upper bounds on $P_t$, 
which have the same time-dependence but slightly differ by numerical factors 
in the definition of the characteristic relaxation time, we show that $P_t$ is a 
stretched-exponential function of time, $\ln(P_t) \sim - t^{1/2}/L^2$. This implies that the distribution function of the first exit time from a fixed interval $[-L,L]$ for such an anomalous diffusion
 has all moments.}

\begin{document}

\maketitle

\section{Introduction} 
Dynamics of a tagged monomer (TM)
of a polymer chain in solution is a practically important physical 
example 
of an
anomalous diffusive process.
While dynamics of the whole chain is dominated by the motion of its center of mass 
which moves diffusively, 
dynamics of a TM 
is coupled to dynamics of
other monomers; as time evolves, 
progressively 
more and more other monomers start to impede dynamics of the TM, 
which ultimately results 
in a subdiffusive motion. 
Such subdiffusive motion
persists up to a certain characteristic time, 
proportional to some
power of the chain length; at greater times 
conventional diffusive behavior is established \cite{doi}.

For the so-called Rouse model of a polymer chain \cite{rouse}, in which a 
chain is considered as a series of $K$ beads linearly 
connected by harmonic springs, 
the TM mean-square displacement $\overline{X^2(t)}$ 
can be calculated exactly. One finds an anomalous diffusion 
law $\overline{X^2(t)} \sim \sqrt{t}$ for times less than the so-called Rouse relaxation 
time $T_R \sim K^2$, (the time needed for some perturbation, i.e. a kink, 
 to spread diffusively along the 
whole chain), and conventional diffusive motion with reduced, by factor $K$, diffusion 
coefficient for times larger than $T_R$.
In fact, for the Rouse model many important dynamical properties
can 
be calculated exactly, e.g., 
the TM position probability distribution function, the 
dynamical structure factor  \cite{doi}, as well as the measure of different 
trajectories of a tagged monomer \cite{b}.  
One may even determine exactly the dynamics of 
the TM of a Rouse chain in 
more complex situations - in random layered flows \cite{osh,maj} or in situations appropriate
to electrophoresis of polyampholytes, i.e., polymers 
whose monomers may be positively or negatively charged and the chain is subject 
to external electric field \cite{osh2}.

Recently, 
following a 
general interest 
in understanding 
subdiffusive and superdiffusive motion,  
a different aspect of anomalous diffusion of 
tagged monomers 
of a Rouse polymer chain has attracted some attention. 
Namely, 
dynamics of 
a Rouse chain in presence of traps or absorbing boundaries reacting with 
some or just one of its monomers
has been analysed \cite{osh3,nech,kardar}. Physically, such a situation is realized 
for polymers diffusing on solid surfaces containing chemically active sites, which may 
react reversibly or irreversibly with any or some of the chain 
monomers temporarily or completely anchoring the chain. 
Conceptually, this question is interesting in its own right
since the answer contains a solution, for anomalous diffusion, of a certain 
first-passage 
time problem, whose general understanding is a basic  
aspect of stochastic processes \cite{sid}. 

Dynamics
of a TM of an infinitely long Rouse
chain in one-dimensional systems in
presence of two absorbing boundaries has been analysed in
Ref.\cite{nech} within a path-integral formalism 
with an exact measure
of trajectories of such a monomer \cite{b}, and a suitably extended classic
method of images. It was shown that the probability $P_t$ 
that the tagged 
monomer commencing 
its motion at the origin
does not escape, during time $t$, from the interval
$[-L,L]$, or, in other words, that it "survives" up to time $t$ in 
presence of two traps placed at $x = -L$ and $x = L$,
obeys
\begin{equation}
\label{b}  \ln\left(P_t\right) \sim - \frac{t^{1/2}}{L^2},
\end{equation} 
i.e., the decay of $P_t$ 
is described by a stretched-exponential function of time. The law in Eq.(\ref{b}) has been 
previously conjectured using heuristic arguments in Ref.\cite{osh3}.

On the other hand, dynamics
 of a tagged monomer of a finite Rouse
chain in one-dimension in presence of absorbing boundaries
 has been analysed numerically in Ref.\cite{kardar}. Here it was
 claimed that $P_t$ is an exponential function of time
\begin{equation}
\label{c} \ln\left(P_t\right) \sim - t,
\end{equation}
and thus decays at a faster rate.
Consequently, the results of Refs.\cite{osh3,nech} and \cite{kardar} 
are in an apparent contradiction with each other.

In this Letter we aim to resolve this controversy by deriving, using the approach outlined in Ref.\cite{i}, rigorous 
lower and upper bounds on the suvival probability $P_t$ of a Rouse chain monomer
in a one-dimensional system with two absorbing bondaries. 
We set out to show here that $P_t$ obeys the following double-sided inequality
\begin{equation}
\label{final}
\frac{1}{4} \leq \frac{- \ln\left(P_t\right)}{t^{1/2}} \left(\frac{4 L^2}{\pi^{3/2}}\right) \leq 1,
\end{equation}
which defines the decay law up to a numerical factor in the characteristic relaxation time.
This inequality confirms the result in Eq.(\ref{b}) and rules out the result in Eq.(\ref{c}).

This paper is outlined as follows: In section 2 we present the notations and write down basic equations. In section 3 we present the results of a heuristic approach, in which the survival probability of a tagged Rouse chain monomer in presence of two absorbing boundaries at $x = \pm L$ is interpreted as the survival probability of a Brownian motion in presence of absorbing boundaries which move away from the origin as $\pm L t^{1/4}$.
Next, in section 4 we derive rigorous lower and upper bounds on $P_t$, which lead to the inequality in Eq.(\ref{final}).
Finally, in section 5, we conclude with a brief recapitulation of our results and discussion.
 
\section{Notations and basic equations}

Dynamics of a discrete Rouse chain comprising  
an infinite number of
monomers is described by a set of Langevin equations \cite{doi}: 
\begin{equation}
\label{lan}
\frac{d X_n(t)}{d t} = \frac{1}{2} \left(X_{n+1}(t) + X_{n-1}(t) - 2 X_n(t)\right) + \zeta_t^{(n)},
\end{equation} 
$X_n(t)$ being an instantaneous position of the $n$-th bead, $- \infty < n < \infty$, and $\zeta_t^{(n)}$ - independent Gaussian white-noise processes, such that
\begin{equation}
\label{jjj}
\overline{\zeta_t^{(n)}} = 0, \;\;\; \overline{\zeta_t^{(n)}
\zeta_{t'}^{(m)}} = \delta_{n,m} \delta(t - t').
\end{equation} 
In Eq.(\ref{jjj}), the bar denotes averaging over thermal histories,
$\delta_{n,m}$ is the Kronecker symbol and $\delta(t)$ is the
delta-function.
Note that, for simplicity of presentation, we have set 
in Eq.(\ref{lan}) the friction constant equal to $1$, the 
spring constant and the temperature equal to $1/2$. These parameters can be easily 
restored in our final results. 

Note, as well,  that Eq.(\ref{lan}) describes the time 
evolution of local heights of 
the Edwards-Wilkinson interface \cite{edw} in one dimension, or the time evolution 
of the difference of local concentrations of $A$ and $B$ species
for diffusion-limited $A+B \to inert$ reactions with random, 
steady, uncorrelated input of $A$ and $B$ \cite{ob}. Our results will thus 
apply to these systems too.

Supposing
that initially all monomers are at the origin, we have that $X_{n=0}(t)$ - position 
of the zeroth TM of an infinitely long Rouse  chain 
at time $t$ - for a given realization of noises
$\zeta_t^{(n)}$ is determined as a portfolio of independent Gaussian processes: 
\begin{equation}
\label{def} X_{n=0}(t) \equiv X(t) = \sum_{n = - \infty}^{\infty} \int_0^t \, d\tau \, \zeta_{\tau}^{(n)} \,  e^{-(t - \tau)}\,  I_n\left(t - \tau\right),
\end{equation}
where $I_n(t-\tau)$ is the modified Bessel function
of order $n$. Eq.(\ref{def}) yields the following expression for the
mean-square displacement of the zeroth monomer of an infinitely long Rouse chain: 
\begin{equation}
\label{msd} \overline{X^2(t)} = t e^{-  2t} \left[I_0(2 t) +
I_1(2 t)\right] = \sqrt{\frac{t}{\pi}} + o\left(\sqrt{t}\right).
\end{equation}
As a matter of fact, any other initial condition can be considered. However, 
the effect of the initial state of the chain on the process in Eq.(\ref{def}) fades out 
quite rapidly; it was observed in Ref.\cite{kardar} that the difference in dynamics 
between 
the case when initially all monomers are at the origin or when 
one starts from an equilibrated configuration 
is rather small. Thus we have chosen the simplest case when $X_n(t=0)=0$ for any $n$.

We note next that since we are concerned
with the large-$t$ behavior, it will not matter much how we define
$\zeta_t^{(n)}$ - as continuous in time functions or as discrete
processes, provided that we keep all essential features of noise. We
thus divide, at fixed $t$, the interval $[0,t]$ into $N$ ($N \gg
1$) small subintervals $\Delta$, (such that $\Delta N \equiv t$),
and assume that $\zeta_t^{(n)}$ is constant and equal to
$\zeta_k^{(n)}/\sqrt{\Delta}$ within the $k$-th subinterval, $k=0,1,
\ldots, N-1$. We suppose that $\{\zeta_k^{(n)}\}$ is an infinite set
of independent random variables with normal distribution $N[0,1]$.

Then, $X(t)$ can be written down as a weighted
sum of an infinite number of independent discrete noise processes:
\begin{eqnarray}
\label{def2} X(t) = \sum_{n = - \infty}^{\infty} \sum_{l=1}^{N} \sigma^{(n)}_{l} \zeta^{(n)}_{N - l},
\end{eqnarray}
with $l$ and $n$-dependent weights
\begin{equation}
\sigma^{(n)}_{l} = \frac{1}{\sqrt{\Delta}} \int_{\Delta (l -
1)}^{\Delta l} du \, e^{- u} \, I_n(u).
\end{equation}
At this point, it is also expedient to introduce 
another property - an effective time-dependent variance $\tilde{\sigma}_l^2$ -
which will emerge in what follows as the key parameter.  
Squaring Eq.(\ref{def2}) and
averaging the resulting expression with respect to distributions
of i.i.d. variables $\{\zeta_k^{(n)}\}$, we get 
\begin{equation}
\label{msd2} \overline{X^2(t)} = \sum_{l=1}^N
\tilde{\sigma}^2_l,
\end{equation}
where the effective variance $\tilde{\sigma}^2_l$ is given by
\begin{eqnarray}
\label{var} \tilde{\sigma}^2_l = \sum_{n = - \infty}^{\infty}
\left(\sigma^{(n)}_l\right)^2 = \nonumber\\
= \frac{1}{\Delta} \int^{\Delta
l}_{\Delta (l-1)} du_1 \int^{\Delta l}_{\Delta (l-1)} du_2 e^{-u_1 -
u_2} \, I_0\left(u_1 + u_2\right).
\end{eqnarray}
The integrations in Eq.(\ref{var}) can be performed exactly,  but the
resulting expression - a combination on nine modified Bessel
functions - is rather cumbersome and is of a little use. In fact,
all information we need to know about $\tilde{\sigma}^2_l$ can
be extracted directly from Eq.(\ref{var}):\\
(a) $\tilde{\sigma}^2_l$ is a \textit{monotonically decreasing}
function of $l$. To see this, it suffice to notice that $exp(-x)
I_0(x)$ is a monotonically decreasing function of $x$; then, one
finds from Eq.(\ref{var}) that $\tilde{\sigma}^2_l$ obeys the
following double-sided inequality:
\begin{equation}
\label{in} \Delta e^{-2 \Delta l} I_0\left(2 \Delta l\right)
\leq \tilde{\sigma}^2_l \leq \Delta e^{-2 \Delta (l - 1)} I_0\left(2
\Delta (l - 1)\right),
\end{equation}
i.e., is bounded from both sides by
monotonically
decreasing
functions of $l$.\\
(b) $\tilde{\sigma}^2_l$ decays as $1/\sqrt{l}$ when $l \to \infty$.
In fact, bounds in Eq.(\ref{in}) become very sharp for $l \gg 1$
and
\begin{equation}
\label{sig} \tilde{\sigma}^2_l \to \Delta e^{-2 \Delta l}
I_0\left(2 \Delta l\right) = \frac{1}{2} \left(\frac{\Delta}{\pi
l}\right)^{1/2} + o\left(\frac{1}{\sqrt{l}}\right)
\end{equation}
Inserting the latter expression into Eq.(\ref{msd2}) and performing summation, we
recover the result in Eq.(\ref{msd}).

Define now the following event: An $N$-step trajectory
$X(t)$, Eq.(\ref{def2}), commencing at the origin,
does not leave the interval $[-L,L]$, $L > 0$, or, in other words, that  ${\rm max}|X(t)| \leq L$.
Such an event takes place, clearly, when the absolute value of 
any ascending partial sum
\begin{equation}
X_{k} = \sum_{n = - \infty}^{\infty} \sum_{l = N - k + 1}^{N} \sigma_l^{(n)} \zeta_{N - l}^{(n)},
\end{equation}
which define positions of the tagged monomer at consecutive discrete "time"
moments $k$, $k=1,2, \ldots, N$, is bounded from above by $L$. 

However, in order to get a convenient "direction" of time in our final results, we will
prefer to work with \textit{descending} partial sums
\begin{equation}
\label{aa}
Y_{k} = \sum_{n = - \infty}^{\infty} \sum_{l = 1}^{k} \sigma_l^{(n)} \zeta_{N - l}^{(n)}.
\end{equation}
Since, evidently,
\begin{equation}
Y_{N - k} + X_{k} \equiv X_{N},
\end{equation}
for any $k$, the trajectory $\{Y_{k}\}$ is exactly the trajectory $\{X_{k}\}$, with the only difference that it evolves in the inverse time $N - k$ and is shifted by a constant (realization-dependent) value $X_{N}$. 
Consequently, the survival probability $P_t = P({\rm max}|X(t)| \leq L)= P({\rm max}|Y(t)| \leq L)$.

To calculate $P({\rm max}|Y(t)| \leq L)$, we proceed as follows. Let $I( {\rm max}|Y(t)| \leq L)$
be the following indicator function:
\begin{equation}
\label{indicator1}
I\Big( {\rm max}|Y(t)| \leq L\Big) =
\begin{cases}
1\,,~~{\rm max}|Y(t)| \leq L\,,\\
0 \,,~~{\rm max}|Y(t)| > L\,.
\end{cases}
\end{equation}
In terms of descending partial sums Eq.(\ref{indicator1}) can be rewritten as
\begin{equation} 
\label{indicator3}
I\Big( {\rm max}|Y(t)| \leq L\Big) = \prod_{k=1}^N
I\Big(|Y_{k}| \leq L\Big).
\end{equation}
Next, let  ${\rm rect}_L(x)$ be a rectangular
function, such that: 
\begin{equation} 
{\rm rect}_L(x) = 
\begin{cases}
1, & ~~ |x| < L,\\
1/2, & ~~ x = \pm L, \\
0,  & ~~ |x| > L.
\end{cases}
\label{indicator}
\end{equation} 
Representing ${\rm rect}_L(x)$ via its Fourier transform:
\begin{equation}
{\rm rect}_L(x) =  \int_{-\infty}^{\infty} \frac{d y}{\pi}
\frac{\sin(L y)}{y} \exp\left[i y x\right],
\end{equation} 
we write down the indicator function
in Eq.({\ref{indicator3}) as the following $N$-fold integral:
\begin{eqnarray} \label{indicator4}
I\Big( {\rm max}|Y(t)| \leq L\Big) &=& \int_{-\infty}^{\infty}
\ldots \int_{-\infty}^{\infty} \prod_{k=1}^N \frac{dy_k}{\pi} 
\frac{\sin(L y_k )}{y_k} \times \nonumber\\
&\times& \exp\left[i \; \sum_{k=1}^N y_k \; Y_{k}\right].
\end{eqnarray} 
Now, in order to determine the desired probability $P_t$, we have to
average the indicator function in Eq.(\ref{indicator4}) with respect
to distributions of i.i.d. variables $\zeta_k^{(n)}$. To do
this, we first rewrite the sum in the exponential in
Eq.(\ref{indicator4}) in the following form 
\begin{equation}
\sum_{k = 1}^N y_k \; Y_{k} = \sum_{n = - \infty}^{\infty}
\sum_{k = 1}^{N} \zeta^{(n)}_{N-k}
\left(\sigma_{k}^{(n)} \sum_{m=k}^N y_m\right).
\end{equation}
Inserting the latter expression into Eq.(\ref{indicator4}) and 
performing averaging, we find that  $P_t$  is given by
\begin{eqnarray} \label{survival}
P_t &=& \int_{-\infty}^{\infty} \ldots \int_{-\infty}^{\infty}
\prod_{k=1}^N \frac{dy_k}{\pi} \; \frac{\sin(L y_k )}{y_k} \times \nonumber\\
&\times& \exp\left[ - \frac{1}{2} \sum_{k = 1}^{N} \tilde{\sigma}^2_{k}
\left(\sum_{m = k}^N y_m\right)^2\right],
\end{eqnarray} 
where the effective variance $\tilde{\sigma}^2_{k}$ has been defined
in Eq.(\ref{var}).

Next, changing the integration variables:
\begin{eqnarray}
Y_1 &=& y_1 + y_2 + \ldots + y_N, \nonumber\\
Y_2 &=& y_2 + y_3 + \ldots + y_{N}, \nonumber\\
Y_3 &=& y_3 + y_4 + \ldots + y_{N}, \nonumber\\
\ldots \nonumber\\
Y_N &=& y_N,
\end{eqnarray}
we obtain
\begin{eqnarray}
\label{u} P_t &=&  \int_{-\infty}^{\infty} \ldots
\int_{-\infty}^{\infty} \prod_{k=1}^{N} \frac{dY_k}{\pi} \;
\frac{\sin\left(L (Y_k - Y_{k+1}) \right)}{Y_k - Y_{k+1}} \times \nonumber\\
&\times& \exp\left[- \frac{1}{2} \tilde{\sigma}_k^2 Y_k^2\right], \; Y_{N+1}
\equiv 0.
\end{eqnarray}
Now, it is expedient to use the following integral identity for the
sinc-function: 
\begin{equation}
\label{uu} \frac{\sin\left(L \left(Y_k - Y_{k+1}\right)
\right)}{Y_k - Y_{k+1}} = \frac{1}{2} \int_{-L}^L dX_k \exp\left[i
X_k \left(Y_k - Y_{k+1}\right)\right].
\end{equation} 
Plugging Eq.(\ref{uu}) into Eq.(\ref{u}), and performing
integrations over $Y_k$-s, we finally arrive at the following
meaningful representation of the survival probability: 
\begin{eqnarray}
\label{f} P_t = \int_{-L}^{L} \ldots \int_{-L}^{L}
\prod_{k=1}^{N} \frac{dX_k}{\sqrt{2 \pi} \tilde{\sigma}_k} \;
\exp\left[- \frac{\left(X_k - X_{k-1}\right)^2}{2 \tilde{\sigma}_k^2}
\right], 
\end{eqnarray}
where $X_0 \equiv 0.$
The integrand in Eq.(\ref{f}) represents the measure of
trajectories of the zeroth monomer of an infinite discrete Rouse
chain. Its continuous-space counterpart has been calculated
previously in Ref.\cite{b}.

\section{Heuristic estimate of $P_t$}

Changing the integration variables, $X_k = \tilde{\sigma}_k x_k$, we can cast 
Eq.(\ref{f}) into the following form:
\begin{eqnarray}
\label{fu} P_t &=& \int_{-L/\tilde{\sigma}_1}^{L/\tilde{\sigma}_1} \int_{-L/\tilde{\sigma}_2}^{L/\tilde{\sigma}_2} \ldots 
\int_{-L/\tilde{\sigma}_N}^{L/\tilde{\sigma}_N} \prod_{k=1}^{N} \frac{dx_k}{\sqrt{2 \pi}} \times \nonumber\\
&\times& \exp\left[- \frac{1}{2} \left(x_k - \frac{\tilde{\sigma}_{k-1}}{\tilde{\sigma}_k} x_{k-1}\right)
\right], \, x_0 \equiv 0.
\end{eqnarray}
Further on, since 
\begin{equation}
\frac{\tilde{\sigma}_{k-1}}{\tilde{\sigma}_k} = 1 + O\left(\frac{1}{k}\right),
\end{equation}
we may expect that, for sufficiently large $k$, this factor will not matter and
\begin{equation}
\left(x_k - \frac{\tilde{\sigma}_{k-1}}{\tilde{\sigma}_k} x_{k-1}\right)^2 \approx  \left(x_k -  x_{k-1}\right)^2.
\end{equation}
Hence, Eq.(\ref{fu}) can be approximated by
\begin{eqnarray}
\label{fuf} P_t &\approx& \int_{-L/\tilde{\sigma}_1}^{L/\tilde{\sigma}_1} \int_{-L/\tilde{\sigma}_2}^{L/\tilde{\sigma}_2} \ldots 
\int_{-L/\tilde{\sigma}_N}^{L/\tilde{\sigma}_N}
\prod_{k=1}^{N} \frac{dx_k}{\sqrt{2 \pi}} \times \nonumber\\
&\times& \exp\left[- \frac{1}{2} \left(x_k - x_{k-1}\right)^2
\right], \, x_0 \equiv 0.
\end{eqnarray}
One notices that the right-hand-side of Eq.(\ref{fuf}) determines the probability that an $N$-step Brownian motion trajectory  does not leave an interval whose boundaries move deterministically away 
from the origin as $\pm L/\tilde{\sigma}_{k}$, i.e., $P_t$ can be approximately defined as 
\begin{equation}
\label{fufu}
P_t \approx \int_{-L/\tilde{\sigma}_t}^{L/\tilde{\sigma}_t} P(X,t) dX,
\end{equation}
where $\tilde{\sigma}_t \sim (\Delta^2/4 \pi t)^{1/4}$, Eq.(\ref{sig}), while $P(X,t)$ obeys
\begin{equation}
\frac{\partial P(X,t)}{\partial t} = \frac{1}{2 \Delta} \frac{\partial^2 P(X,t)}{\partial X^2}, \, P(X, t = 0) = \delta(X),
\end{equation}
and
\begin{equation}
P(X = \pm L/\tilde{\sigma}_t,t) = 0.
\end{equation}
We estimate next $P_t$ defined by Eq.(\ref{fufu}) using an adiabatic 
approximation discussed in Ref.\cite{kr}.  The basic idea behind this approximation is that, if 
the boundary advances sufficiently slowly, the density distribution approaches the same form as in the \textit{fixed} boundary case, except that the parameters in this probability distribution acquire time dependence to satisfy moving boundary conditions \cite{kr}. We find that, within 
this  approximation, in the leading order
\begin{equation}
P(X,t) \approx \exp\left[- \frac{\pi^2}{8 \Delta} \int^t_0 \left(\frac{\tilde{\sigma}_{\tau}}{L}\right)^2 d\tau\right] \cos\left(\frac{\pi X \tilde{\sigma}_t}{2 L}\right),
\end{equation}
which yields the following estimate:
\begin{equation}
P_t \approx \exp\left[- \frac{\pi^2}{8 L^2} \left(\frac{t}{\pi}\right)^{1/2}\right].
\end{equation}
Note that this estimate  agrees with Eq.(\ref{b}).

\section{Rigorous bounds on the survival probability $P_t$}

The $N$-fold integral in Eq.(\ref{f}) can not be, of course, performed 
exactly and recourse has to be made to controllable approximations. 
Below we construct rigorous lower and upper bounds on $P_t$ in Eq.(\ref{f}), 
which both have the same time dependence
defining in such a way an asymptotically exact (up to a constant factors in the characteristic relaxation time) result.

The method we use here is based on the approach outlined in Ref.\cite{i} within the context 
of the Riemann-Liouville fractional Brownian motion. In constructing bounds, we will take advantage of the following two facts:\\
(a) The effective dispersion $\tilde{\sigma}_k$ in Eq.(\ref{f})
is a monotonically decreasing function of time.\\
(b) A fundamental property of $P_t$ defined by Eq.(\ref{f}) is that $P_t = P_t(\tilde{\sigma}_1,\tilde{\sigma}_2,\tilde{\sigma}_3, \ldots,\tilde{\sigma}_N)$ is a monotonically decreasing function of any variable $\tilde{\sigma}_k$ \cite{i}. 

This signifies that replacing any or all $\tilde{\sigma}_k$ by $\Sigma(k)$, such that
$\tilde{\sigma}_k \leq \Sigma(k)$, we will decrease the survival
probability and arrive at the \textit{lower} bound on $P_t$; if, on
contrary, we will replace one or all $\tilde{\sigma}_k$ by
$\tilde{\Sigma}(k)$, such that $\tilde{\sigma}_k \geq \tilde{\Sigma}(k)$, we will \textit{increase} the survival probability and
obtain an \textit{upper} bound on $P_t$.

We start with an upper bound on $P_t$. Since $\tilde{\sigma}_k$ is a monotonically decreasing function of $k$, we have that
\begin{equation}
\tilde{\sigma}_k \geq \tilde{\Sigma}(k) \equiv \tilde{\sigma}_N, 
\end{equation}
which holds for any $k$. The equality is attained only for $k = N$.

Hence, $P_t$ in Eq.(\ref{f}) is bounded from \textit{above} by
\begin{eqnarray}
\label{ff} P_t \leq \int_{-L}^{L} \ldots \int_{-L}^{L}
\prod_{k=1}^{N} \frac{dX_k}{\sqrt{2 \pi} \tilde{\sigma}_N} \;
\exp\left[- \frac{\left(X_k - X_{k-1}\right)^2}{2 \tilde{\sigma}_N^2}
\right] = \nonumber\\
=  \int_{-L/\tilde{\sigma}_N}^{L/\tilde{\sigma}_N} \ldots \int_{-L/\tilde{\sigma}_N}^{L/\tilde{\sigma}_N}
\prod_{k=1}^{N} \frac{dx_k}{\sqrt{2 \pi} } \;
\exp\left[- \frac{\left(x_k - x_{k-1}\right)^2}{2}
\right]
\end{eqnarray}
where $X_0 \equiv 0$ and $x_0 \equiv 0$.

Expression in the second line in Eq.(\ref{ff}) 
determines the probability that an $N$-step Brownian motion 
trajectory does not leave an interval with fixed boundaries $\pm L/\tilde{\sigma}_N$, which is a classic problem in the probability theory (see, e.g., Ref.\cite{feller}). Consequently, we find that at sufficiently large times
\begin{equation}
\label{gg}
P_t \leq \exp\left[- \frac{\pi^2}{8} \frac{\tilde{\sigma}^2_N}{L^2} N\right] = \exp\left[- \frac{\pi^2}{16 L^2} \left(\frac{t}{\pi}\right)^{1/2}\right].
\end{equation}

Consider now a lower bound on $P_t$ in Eq.(\ref{f}). To construct such a bound, we turn back to the definition of $\tilde{\sigma}_k$ in Eq.(\ref{var}). Recollecting that $exp(-x) I_0(x)$ is a monotonically decreasing function of $x$, we have
that for any $u_1,u_2 \geq 0$,
\begin{equation}
e^{- u_1 - u_2} I_0\left(u_1 + u_2\right) \leq e^{- u_2} I_0\left(u_2\right).
\end{equation}
Consequently, $\tilde{\sigma}_k^2$ is bounded as
\begin{equation}
\tilde{\sigma}_k^2 \leq  \int^{\Delta k}_{\Delta (k-1)} du_2 e^{-
u_2} \, I_0\left(u_2\right).
\end{equation}
Further on, since for any $u_2 > 0$,
\begin{equation}
e^{-
u_2} \, I_0\left(u_2\right) \leq \frac{1}{\sqrt{\pi u_2}},
\end{equation}
we have that the following bound holds:
\begin{equation}
\tilde{\sigma}_k^2 \leq \Sigma^2(k) = 2 \left(\frac{\Delta}{\pi}\right)^{1/2} \left(T_k - T_{k - 1}\right), 
\end{equation}
where
\begin{equation}
\label{time}
T_k = k^{1/2}.
\end{equation}
Hence, the survival probability $P_t$ in Eq.(\ref{f}) is bounded from \textit{below} by
\begin{eqnarray}
\label{fif} &P_t& \geq \int_{-L}^{L} \ldots \int_{-L}^{L}
\prod_{k=1}^{N} \frac{dX_k}{\sqrt{2 \pi} \Sigma(k)} \;
\exp\left[- \frac{\left(X_k - X_{k-1}\right)^2}{2 \Sigma^2(k)}
\right]  \nonumber\\
&=&  \int_{-L (\frac{\pi}{4 \Delta})^{1/4}}^{L(\frac{\pi}{4 \Delta})^{1/4}} \ldots \int_{-L(\frac{\pi}{4 \Delta})^{1/4}}^{L(\frac{\pi}{4 \Delta})^{1/4}}
\prod_{k=1}^{N} \frac{dx_k}{\sqrt{2 \pi \left(T_k - T_{k-1}\right)} }  \nonumber\\
&\times& \exp\left[- \frac{\left(x_k - x_{k-1}\right)^2}{2 \left(T_k - T_{k - 1}\right)}
\right], \, X_0 \equiv 0, \, x_0 \equiv 0.
\end{eqnarray}

We notice now that the expression in the second line in Eq.(\ref{fif}) 
defines the probability that an $N$-step trajectory of Brownian 
motion, evolving in time $T_t$, Eq.(\ref{time}), does not leave the 
interval $[-L (\frac{\pi}{4 \Delta})^{1/4},L (\frac{\pi}{4 \Delta})^{1/4}]$. 
Hence, $P_t$ in Eq.(\ref{f}) is bounded from below by
\begin{equation}
\label{jj}
P_t \geq  \exp\left[- \frac{\pi^2}{8 L^2} \left(\frac{4 \Delta}{\pi}\right)^{1/2} T_N\right] = \exp\left[- \frac{\pi^2}{4 L^2} \left(\frac{t}{\pi}\right)^{1/2}\right].
\end{equation}
This bound rules out an exponentially fast decay of the survival probability suggested in Ref.\cite{kardar}.

Finally, combining the lower and the upper bounds, we obtain the double-sided 
inequality obeyed by $P_t$, Eq.(\ref{final}),
which defines the decay of the survival probability $P_t$ up to 
a numerical factor in the characteristic relaxation time.

\section{Conclusions}

To conclude, we have studied  the long-time asymptotical behavior of the
probability $P_t$
that a tagged monomer of an infinitely long Rouse 
chain will not escape, up to time $t$, from an interval $[-L,L]$. We have shown, 
by constructing rigorous lower and upper bounds on $P_t$, which both have the same dependence on time but slightly differ, by numerical factors, in the definition of the characteristic relaxation time, that $P_t$ follows $\ln(P_t) \sim - t^{1/2}/L^2$. This decay law confirms our earlier predictions based on uncontrollable approaches \cite{osh3,nech} and contradicts a recent prediction $\ln(P_t) \sim - t$ based on numerical simulations \cite{kardar}.

This result implies that the probability distribution 
function of the first-exit time from an interval $[-L,L]$ for the anomalous diffusion process executed by the tagged monomer of an infinitely long Rouse chain has a stretched-exponential tail $\sim \exp(- t^{1/2}/L^2)$ and thus has \textit{all} moments.    

We note that the obtained decay law agrees, as well, with the 
general result on the survival probability of the 
Riemann-Liouville fractional Brownian motion (fBm) in presence of absorbing boundaries
\cite{i} when the Hurst index $H = 1/4$. To realize that this is not a coincidence and the 
dynamics of a tagged monomer of an infinitely long Rouse chain is indeed in the fBm "universality class", consider Eq.(\ref{lan}), in which, for simplicity, we treat $n$ as a continuous variable. Then, we have that $X_{n=0}(t)$ obeys
\begin{equation}
\label{ww}
X_{n=0}(t) = \int^t_0 \frac{\zeta_{\tau} d \tau}{\left(t - \tau\right)^{1/2}} 
\int_{-\infty}^{\infty} dn  \exp\left[- \frac{n^2}{2 (t - \tau)}\right] f_n,   
\end{equation}  
where $\zeta_t$ is a white noise in time and $f_n$ is a white noise of 
variable $n$, $-\infty < n < \infty$.
Notice now that $\exp[- n^2/2 (t - \tau)]$ is a bell-shaped function which broadens with time, which signifies that as time progresses, more and more monomers start to affect the dynamics of the zeroth monomer. Let us replace, for simplicity, $\exp[- n^2/2 (t - \tau)]$ by a rectangular function ${\rm rect}(n)$, such that ${\rm rect}(n) = 1$ for $n \leq \sqrt{t - \tau}$ and ${\rm rect}(n) \equiv 0 $ for $n >  \sqrt{t - \tau}$. Then, Eq.(\ref{ww}) reads
\begin{equation}
X_{n=0}(t) \approx \int^t_0 \frac{\zeta_{\tau} d \tau}{\left(t - \tau\right)^{1/2}} 
\int_{-\sqrt{t - \tau}}^{\sqrt{t - \tau}} dn  f_n .  
\end{equation}
Now, one expects that, for typical realizations of $f_n$,
\begin{equation}
 \int_{-A}^{A} dn  f_n \sim A^{1/2},   
\end{equation}
and hence, $X_{n=0}(t)$ follows
\begin{equation}
X_{n=0}(t) \approx \int^t_0 \frac{\zeta_{\tau} d \tau}{\left(t - \tau\right)^{1/4}}, 
\end{equation}
which is exactly the Riemann-Liouville fractional Brownian motion with Hurst index $H = 1/4$
studied in Ref.\cite{i}. 

\acknowledgments

The author acknowledges helpful discussions with M.Kardar, J.Klafter
and S.Nechaev.

\end{document}